\documentclass[aps,prl,twocolumn]{revtex4}

\usepackage[english]{babel}
\usepackage{amssymb,latexsym,amsmath}
\usepackage{graphicx}

\begin{document}

\title{SU(2/1) gauge-Higgs unification}
\author{E.K. Loginov}
\email{ek.loginov@mail.ru}
\affiliation{Ivanovo State University, 153025, Russia}

\begin{abstract}
We discuss a question whether the observed Weinberg angle and Higgs mass are calculable in the formalism based on a construction in which the electroweak gauge group $SU(2)\times U(1)_{Y}$ is embedded in the graded Lie group $SU(2/1)$. Here we follow original works of Ne'eman and Fairlie believing that bosonic fields take their values in the Lie superalgebra and  fermionic fields take their values in its representation space. At the same time, our approach differs significantly. The main one is that while for them the gauge symmetry group is $SU(2/1)$, here we consider only symmetries generated by its even subgroup, i.e., symmetries of the standard electroweak model. The reason is that such formalism fixes the quartic Higgs coupling and at the same time removes the sign and statistics problems. The main result is that the presented model predicts values of the Weinberg angle and the Higgs mass correctly up to the two-loop level. Moreover, the model sets the unification scale coinciding with the electroweak scale and automatically describes the fermions correctly with the correct quark and lepton charges.
\end{abstract}
\maketitle

\section{Introduction}

The standard model (SM) is a mathematically consistent renormalizable field theory which consistent with all experimental facts. It successfully predicted the existence of the weak neutral current and masses of the $W$ and $Z$ bosons, and the charm quark, as necessitated by the GIM mechanism. At the same time some values of the SM parameters are not calculable in the theory, notably, the fermion mass hierarchy, the hierarchy of symmetry-breaking scales, and the Higgs boson mass. Furthermore, the description of electroweak symmetry breaking with a Higgs boson suffers from several instabilities at the quantum level. The SM particles give unnaturally large corrections to the Higgs mass. They destabilize the Higgs vacuum expectation value and tend to push it towards the ultraviolet cutoff of the SM.
\par
One way to protect the divergence is to identify the scalar as the extra components of some
gauge field in extra dimensions. This idea has been exploited for a long time~\cite{mant79,hoso83,hata98}, but only recently within the context of orbifolds it has been implemented in realistic models ~\cite{csak01,dval02,hall02,csak03,burd03,scru03,medi07,care07,hoso08} (see also the
review~\cite{groj07}). Another early theory which invoked the notion of additional dimensions
was that of~\cite{neem79,fair79} (see also~\cite{neem05}) in which the gauge $SU(3)$ group was
replaced by the graded Lie group $SU(2/1)$. In this formalism, the even (bosonic) part of the
$su(2/1)$ algebra defines the $SU(2)\times U(1)_{Y}$ gauge sectors of the SM, while the Higgs
sector is identified as the odd (fermionic) part of the algebra. Interestingly, the
fundamental representation of $SU(2/1)$ exactly corresponds to the lepton triplet, which
contains an $SU(2)$ doublet and a singlet. Moreover, it was noted~\cite{dond79,neem80} that
$SU(2/1)$ also admits a four-dimensional representation exactly fitting the quarks, with two
right singlets and a left doublet with electric charges $2/3$ and $-1/3$. All this makes this
model very attractive. At the same time, it have a number of phenomenological problems.
\par
First, since the group $SU(2/1)$ is simple, it follows that the model fixes the ratio of the
gauge couplings (and hence the Weinberg angle) and the quartic Higgs coupling (and hence the
Higgs mass). The ratio $g/g'=\sqrt{3}$ and the value $\sin^{2}\theta_{W}=1/4$ have been
already found in the first works~\cite{neem79,fair79}. The tree-level values of the Higgs mass
have been predicted in the works~\cite{fair79,neem82,neem86,neem96}: $M_{H}\simeq 426$, 246,
161 and 130 GeV respectively. Unlike the three first results, the last value was obtained by
the one-loop approximation of the tree-level Higgs mass at the scale $M_{0}=2M_{W}$. However,
as was shown in~\cite{ayde13}, the value $\sin^{2}\theta_{W}=1/4$ select the scale
$M_{0}\simeq 4$ TeV in which the renormalization group running leads to predictions of the
Higgs mass around 170 GeV. Thus, all these predictions are in contradiction with the
the experimental data.
\par
The second difficult is the so-called ``sign problem'', that arises from the fact that in the
$SU(2/1)$ gauge theory the kinetic energy of the vector bosons (which is associated with the
supertrace of the product of generators) is not positive-definite, and hence at least one of
the kinetic energy term is going to have the wrong sign~\cite{eccl80}. Finally, there is a
so-called ``statistics problem'' associated with the fact that the parameters that multiply
the odd generators of $SU(2/1)$ have to be Grassmann numbers, so that the Higgs fields have to
be anticommuting among themselves, in total disagreement with the fundamentals of quantum
field theory. All these problems are very serious and so far are not solved.
\par
In this paper we follow the original works of Ne'eman and Fairlie~\cite{fair79,neem79}
believing that all bosonic fields of the presented model take their values in the superalgebra
Lie $su(2/1)$ while the fermionic fields take their values in a representation space of
$su(2/1)$. At the same time the model contains a number of essential differences from the model of Ne'eman and Fairlie. The main of them is that while for them the gauge symmetry group is $SU(2/1)$, here we admit only symmetries generated by its even subgroup, i.e., symmetries of the standard electroweak model. Despite the fact that this requirement seems some what unnatural (in the section 2, we'll make an attempt to dispel these doubts), it is very attractive as it eliminates the sign and  statistics problems. Indeed, since the trace of any element of the superalgebra is $SU(2)\times U(1)$ gauge-invariant, it follows that the kinetic energy of the vector bosons (which is associated with the trace) is positive-definite, and hence the sign problem is removed. Further, since the adjoint action of $SU(2)\times U(1)$ on $su(2/1)$ does not contain anticommutators, it follows that elements of the Grassmann algebra cannot be parameters of the action. Hence the Higgs doublet is not associated with the odd generators, it only defines a representation for the even subgroup of $SU(1/2)$, and consequently has not the wrong statistics.
\par
This paper is organized as follows. In the next section we present a simple model of gauge-Higgs unification that gives a prediction for the tree-level value of the Higgs boson mass. The section 3 contains the basic results. In this section we identify the Higgs potential with the extra component of a five-dimensional vector field and build the $SU(2)\times U(1)$ gauge-invariant Lagrangian. We then find the tree-level values of the Weinberg angle and the Higgs mass. In the next two sections we compute values of these quantities in the one-loop and two-loop approximations. Finally, section 6 is devoted to some general conclusions, and some technical details are collected in the appendix.

\section{Naive construction}

We begin by considering the standard electroweak model with its fermionic sector composed of
the first family leptons only. With the fermions the gauge-invariant Lagrangian takes on the
form
\begin{align}\label{2-06}
\mathcal{L}_{\psi}&=\bar\psi_{L}i\gamma^{\mu}D_{\mu}\psi_{L}+\bar
\psi_{R}i\gamma^{\mu}D_{\mu}\psi_{R}\notag\\
&+f_{e}\left(\bar\psi_{L}\phi\psi_{R}+\bar\psi_{R}\phi^{\dag}\psi_{L}\right),
\end{align}
where the left-handed field $\psi_{L}=(\nu_{L},e_{L})$ transforms as $SU(2)$ doublet and
$\psi_{R}=e_{R}$ is a singlet. We rewrite the Lagrangian (\ref{2-06}) in a slightly different
way. Let
\begin{equation}\label{2-04}
\widetilde D_{\mu}=D_{\mu}+\frac{i}{4}\gamma_{\mu}M,
\end{equation}
where the covariant derivative and the scalar field are defined by
\begin{align}
D_{\mu}&=\partial_{\mu}-i\frac{g}{2}A^{i}_{\mu}\lambda_{i}-i\frac{g'}{2}B_{\mu}\lambda_0,
\label{2-09}\\
M&=m-k(\phi_{4}\lambda_{4}+\phi_{i+4}\lambda_{i+4})\label{2-10}.
\end{align}
Here $g$ and $g'$ are the usual coupling constant, $\lambda_{1},\dots,\lambda_{7}$ are the
standard Gell-Mann matrices and $\lambda_{0}=\text{diag}(-1,-1,-2)$. The constant $m$ and the
normalizing factor $k$ will be determined later. Consider the Lagrangian
\begin{equation}\label{2-07}
\mathcal{L}_{\varPsi}=\overline\varPsi i\gamma^{\mu}\widetilde D_{\mu}\varPsi,
\end{equation}
where the triplet $\varPsi$ is composed of a doublet $\psi_{L}$ and a singlet $\psi_{R}$. It
is easy to see that this Lagrangian is coincided with (\ref{2-06}) as soon as
\begin{equation}\label{2-02}
\phi=\begin{pmatrix}\phi_{4}-i\phi_{5}\\\phi_{6}-i\phi_{7}\end{pmatrix}\quad\text{and}\quad
k=f_{e}.
\end{equation}
\par
In order to clarify the meaning of the operator $\widetilde D_{\mu}$, we consider the
Lagrangian for a free spinor field
\begin{equation}\label{2-08}
\mathcal{L}_{0}=\bar\psi(i\gamma^{\mu}\partial_{\mu}-m)\psi.
\end{equation}
Suppose the field $\psi=\psi_{L}+\psi_{R}$ transforms as $n$-plet under an unitary group $G$.
Then $\mathcal{L}_{0}$ has a global $G$ symmetry. Replacing $\partial_{\mu}\psi$ by the
covariant derivative $D_{\mu}\psi$, we obtain a gauge-invariant Lagrangian. Now let the
left-handed field $\psi_{L}$ transforms as $G$ $n$-plet, whereas the right-handed field
$\psi_{R}$ transforms as $H\subset G$ $m$-plet. If $H\ne G$, then the Lagrangian (\ref{2-08})
is not invariant even under the global $(G,H)$ symmetry. To construct a gauge-invariant
Lagrangian we proceed as follows. We replace $\psi$ by the $(n+m)$-plet
$\varPsi=(\psi_{L},\psi_{R}$), the partial derivative $\partial_{\mu}\psi$ by the covariant
derivative $D_{\mu}\varPsi$, and the constant $m$ by the operator $M$ of the (\ref{2-10}) type. If we now assume that the $n\times m$ matrix $\phi$ is transforms under left action of $G$ and right action of $H$, then we get the gauge-invariant Lagrangian
\begin{equation}\label{2-13}
\mathcal{L}=\overline\varPsi(i\gamma^{\mu}D_{\mu}-M)\varPsi
\end{equation}
which is exactly the same as the Lagrangian (\ref{2-07}) for the case $G=SU(2)\times U(1)$ and
$H=U(1)$. Thus, the expression (\ref{2-04}) can be regarded as a ``deformation'' of the
covariant derivative $D_{\mu}$, which is defined by the choice of a subgroup in the gauge symmetry
group. Note also that the Lagrangian (\ref{2-13}) (as well as (\ref{2-01})) is invariant under the global $U(m+n)$ transformation
\begin{equation}
\varPsi\to U\varPsi\quad\text{and}\quad\widetilde D_{\mu}\to U\widetilde D_{\mu}U^{-1}
\end{equation}
mixing the left- and right-handed fermions and the vector and scalar bosons. This was first noted in~\cite{tera78}.
\par
Now we consider the deformation of the Yang-Mills Lagrangian. Let
\begin{equation}\label{2-01}
\mathcal{L}_{A,\varPhi}=\frac{1}{2g^2}\text{Tr}(\widetilde{F}_{\mu\nu}\widetilde{F}^{\mu\nu}),
\end{equation}
where the field $\widetilde{F}_{\mu\nu}$ is defined by
\begin{equation}\label{2-12}
[\widetilde{D}_{\mu},\widetilde{D}_{\nu}]
=\widetilde{F}_{\mu\nu}-i\frac{g'}{2}(\partial_{\mu}B_{\nu}-\partial_{\nu}B_{\mu})\lambda_{0}.
\end{equation}
With the complex doublet (\ref{2-02}) and its covariant derivative that given by
\begin{equation}\label{2-11}
D_{\mu}\phi=\partial_{\mu}\phi-i\frac{g}{2}A^{a}_{\mu}\sigma_{a}\phi,
\end{equation}
the Lagrangian (\ref{2-01}) takes the form
\begin{align}\label{2-05}
\mathcal{L}_{A,\varPhi}&=-\frac14F^{i}_{\mu\nu}F^{i\mu\nu}\notag\\
&-\frac{3k^{2}}{2g^{2}}\left[(D_{\mu}\phi)^{\dag}D^{\mu}\phi
+3m^{2}\phi^{\dag}\phi+\frac{k^{2}}{2}(\phi^{\dag}\phi)^2\right].
\end{align}
We choose $k^{2}=-g^{2}/3$ and identify $\phi$ with the Higgs doublet. Then the Lagrangian
(\ref{2-05}) is precisely the Weinberg-Salam model Lagrangian (more precisely, it is the
bosonic part of the Lagrangian without the $U(1)$-term), in which the gauge and quartic couplings are related by
\begin{equation}\label{2-03}
\lambda=\frac{g^2}{3}.
\end{equation}
Since the mass of the physical Higgs scalar particle (in the tree-level approximation) is
$M_{H}=\sqrt{2\lambda}v$, where $v$ is the vacuum expectation value, the model gives a prediction for the mass of the Higgs boson.

\section{Gauge-Higgs model}

We have shown in the last section that the SM Lagrangian can be obtained as a deformation of
the Yang-Mills Lagrangian. In this case the quartic Higgs coupling is fixed, and hence the
model gives a theoretical prediction for the Higgs boson mass. However it falls short of being
a quite satisfactory model because the origin of the normalizing constant $k$ is far from
clear, as well as the inclusion of the quarks. In addition, by ignoring the $U(1)$-term in the
bosonic part of the Lagrangian we significantly impoverish the model. In this section
we shall present a solution of the problems.
\par
Let us consider first the commutator (\ref{2-12}) in detail. It is not difficult to shown that
\begin{align}
[\lambda_{a},\lambda_{b}]&=2ic_{abc}\lambda_{c},\label{3-01}\\
[\lambda_{a},\gamma_{\mu}\lambda_{\bar{b}}]
&=2i\gamma_{\mu}f^{\bar{c}}_{a\bar{b}}\lambda_{\bar{c}},\label{3-02}\\
[\gamma_{\mu}\lambda_{\bar{a}},\gamma_{\nu}\lambda_{\bar{b}}]
&=2\gamma_{\mu}\gamma_{\nu}d^{c}_{\bar{a}\bar{b}}\lambda_{c},\label{3-03}
\end{align}
where $\mu\ne\nu$. Comparing this with the formulas (\ref{A-01})--(\ref{A-03}) in Appendix, we
see that the projections $\gamma_{\mu}\lambda_{\bar{a}}\to\lambda_{\bar{a}}$ and
$\gamma_{\mu}\gamma_{\nu}\lambda_{a}\to\lambda_{a}$ transform (\ref{3-01})--(\ref{3-03}) into
the commutation relations of the Lie superalgebra $su(2/1)$. Of course, these two types of commutation relations are totally different. Nevertheless, since we take the trace in (\ref{2-01}) of both the Gell-Mann matrices and the Dirac $\gamma$-matrices, it follows that ignoring the latter only leads to a change of coefficients in the right side of (\ref{2-05}).
With this in mind, we can now turn to the construction of the gauge-Higgs model in which the electroweak gauge group $SU(2)\times U(1)_{Y}$ is embedded in the graded Lie group $SU(2/1)$. To this end, we redefine the covariant
derivative (\ref{2-09}) and the scalar field (\ref{2-10}) by
\begin{align}
D_{\mu}&=\partial_{\mu}-i\frac{g}{2}A_{\mu}^{a}\lambda_{a},\label{3-05}\\
M&=m-\frac{g}{2}\phi_{\bar{a}}\lambda_{\bar{a}},\label{3-06}
\end{align}
where $\lambda_{a}$ and $\lambda_{\bar{a}}$ are $4\times4$ matrices, which are defined in
appendix A. We suppose that the fields contained in the even part of the superalgebra $su(2/1)$ are the Yang-Mills gauge fields while the massless $SU(2)$ doublet contained in the odd part of $su(2/1)$ is the Higgs doublet. Consider the Lagrangian
\begin{equation}\label{3-04}
\mathcal{L}_{A,\varPhi}=\frac{1}{2g^2}\text{Tr}\left(F_{MN}F^{MN}\right),
\end{equation}
where the field $F_{MN}$ is defined by
\begin{equation}\label{3-10}
F_{\mu\nu}=[D_{\mu},D_{\nu}],\quad F_{\mu5}=[D_{\mu},M],\quad F_{55}=\{M,M\}.
\end{equation}
With the complex doublet $\phi$ and its covariant derivative $D_{\mu}\phi$ are given in
(\ref{2-02}) and (\ref{2-11}) the Lagrangian (\ref{3-04}) takes on the form
\begin{align}\label{3-07}
\mathcal{L}_{A,\varPhi}^{(n)}&=-\frac{1}{4}F^{i}_{\mu\nu}F^{i\mu\nu}
-\frac{2+n^{2}}{4n^{2}}F^{0}_{\mu\nu}F^{0\mu\nu}\notag\\
&-\frac{1}{2n}\left[(D_{\mu}\phi)^{\dag}D^{\mu}\phi+3m^{2}\phi^{\dag}\phi
-\frac{1+n^{2}}{4n}g^2(\phi^{\dag}\phi)^2\right].
\end{align}
Note that the definition of coupling constants depends on the normalization of the generators.
Since in the non-Abelian case the normalization of the generators is fixed by the nonlinear
commutation relations, it follows that the generators of $su(2/1)$ must be similarly
normalized. We will return to this point later.
\par
We now consider the calculations in more detail. To this end, we rewrite the second terms on
the right-hand side of (\ref{3-05}) and (\ref{3-06}) in the explicit form
\begin{equation}\label{3-08}
A^{a}_{\mu}\lambda_{a}=\begin{pmatrix}A_{\mu}^{k}\sigma_{k}+\frac{1}{n}A_{\mu}^{0}&0&0\\
0&\frac{n+1}{n}A_{\mu}^{0}&0\\0&0&\frac{1-n}{n}A_{\mu}^{0}\end{pmatrix},
\end{equation}
where $\sigma_{k}$ are the standard Pauli matrices, and
\begin{equation}\label{3-09}
\phi_{\bar{a}}\lambda_{\bar{a}}=\begin{pmatrix}0&-\sqrt{\frac{n+1}{2n}}\,\tilde{\phi}
&\sqrt{\frac{n-1}{2n}}\,\phi\\
\sqrt{\frac{n+1}{2n}}\,\tilde{\phi}^{\dag}&0&0\\
\sqrt{\frac{n-1}{2n}}\,\phi^{\dag}&0&0\end{pmatrix},
\end{equation}
where the isodoublet $\tilde{\phi}=i\sigma_{2}\phi^{*}$. (Note that the third and fourth terms
on the right-hand side of (\ref{3-07}) contain only the combinations $\phi^{\dag}\phi$ because
of the identities $\text{Tr}(\phi\phi^{\dag})=\phi^{\dag}\phi=\tilde{\phi}^{\dag}\tilde{\phi}$
and $\text{Tr}(\phi\phi^{\dag})^{2}=(\phi^{\dag}\phi)^{2}$; the same is true for the second
therm.) In order to give physical meaning of the Lagrangian (\ref{3-07}), we consider the
covariant derivative $D_{\mu}\varPsi$ with the fermions $\psi_{L}$, $\psi_{R_{1}}$ and
$\psi_{R_{2}}$ arranged in a quartet $\varPsi$ transforming as a doublet together with two
singlets in the internal space. It is not difficult to see that the Lagrangian
$\overline{\varPsi}i\gamma^{\mu}D_{\mu}\varPsi$ has physical meaning only if $n=-1$ or $n=3$.
Indeed, it follows from (\ref{3-08}) that only in this case the factors in front of
$A_{\mu}^{0}$ coincide with the fermionic hypercharge values. We call the representations of
$su(2/1)$ with $n=-1$ and $n=3$ the lepton and quark representation respectively.
\par
We now construct a combined quark-lepton representation. We first note that quarks have the
three colours and, therefore, the dimensionality of the quark representation has to be
enlarged. The simplest way to do this is to rewrite the commutation relations
(\ref{A-01})--(\ref{A-03}) in the form
\begin{align}
[\lambda_{a}^{\mu},\lambda_{b}^{\nu}]
&=2i\delta^{\mu}_{\nu}c_{abc}\lambda_{c}^{\nu},\\
[\lambda_{a}^{\mu},\lambda_{\bar{b}}^{\nu}]
&=2i\delta^{\mu}_{\nu}f^{\bar{c}}_{a\bar{b}}\lambda_{\bar{c}}^{\nu},\\
\{\lambda_{\bar{a}}^{\mu},\lambda_{\bar{b}}^{\nu}\}
&=2\delta^{\mu}_{\nu}d^{c}_{\bar{a}\bar{b}}\lambda_{c}^{\nu},
\end{align}
where the metric $g^{\mu\nu}$ has the signature $(+---)$ and no summation over $\nu$. As
above, the matrices $\lambda_{\bar{a}}^{\mu}$ and $\lambda_{a}^{\mu}$ define the
$n$-representations of $su(2/1)$. Suppose $n=-1$ as $\mu=0$ and $n=3$ otherwise. We obtain a
Lie superalgebra with the basis $\{\lambda_{a}^{\mu},\lambda_{\bar{a}}^{\mu}\}$, which is
isomorphic to the direct sum of four superalgebras $su(2/1)$. Its subalgebra with the
generators $\lambda_{a}=\sum_{\mu}\lambda_{a}^{\mu}$ and
$\lambda_{\bar{a}}=\sum_{\mu}\lambda_{\bar{a}}^{\mu}$ give us a representation of $su(2/1)$.
Obviously, this representation can be is realized by $16\times16$ block matrices. We shall
refer to this representation as the quark-lepton representation. Redefining the even
generators as $\tilde{\lambda}_{0}=\sqrt{3/10}\,\lambda_{0}$ and
$\tilde{\lambda}_{k}=\sqrt{1/2}\,\lambda_{k}$, we obtain the following normalization for the
quark-lepton representation:
\begin{equation}\label{3-12}
\text{Tr}\left(\tilde{\lambda}_{a}\tilde{\lambda}_{b}\right)=4\delta_{ab},\quad
\text{Tr}\left(\lambda_{\bar{a}}^{\mu}\lambda_{\bar{b}\mu}\right) =4\delta_{\bar{a}\bar{b}},
\end{equation}
where all generators are similarly normalized. Now we can begin the construction of the
realistic gauge-Higgs model.
\par
Again let us examine the Lagrangian (\ref{3-04}), where we take the generators of the
quark-lepton representation which are normalized by the conditions (\ref{3-12}), and where the
factor $1/2$ is replaced by $1/4$. Using the same procedure as above, we get instead of
(\ref{3-07}) the Lagrangian
\begin{align}\label{4-02}
\mathcal{L}_{A,\varPhi}=&-\frac14F^{a}_{\mu\nu}F^{a\mu\nu}\notag\\
&+\frac{1}{2}\left[(D_{\mu}\phi)^{\dag}D^{\mu}\phi
+3m^2\phi^{\dag}\phi-\frac{g^{2}}{6}(\phi^{\dag}\phi)^2\right].
\end{align}
In order to obtain the standard model Lagrangian we only need to identify $\phi$ with the
Higgs doublet and $A^{k}_{\mu}=\tilde{A}^{k}_{\mu}/\sqrt{2}$ with the $SU(2)_{L}$ gauge
bosons. The last condition is a consequence of the chosen normalization and the obvious
identity $A^{k}_{\mu}\lambda_{k}=\tilde{A}^{k}_{\mu}\tilde{\lambda}_{k}$ (provided that the
gauge coupling is unchanged).
\par
As in (\ref{2-05}), the quartic coupling is given by (\ref{2-03}). In contrast to the model
discussed in the previous section, however, in the Lagrangian (\ref{4-02}) there is no the
arbitrary parameter $k$. Moreover, in this model there is only one gauge coupling constant
$g$. Therefore it also predict a relation between the coupling constants of $SU(2)$ and
$U(1)_{Y}$ at a unification scale $M_{0}$. Indeed, the definition of coupling constants
depends on the normalization of the generators. For the superalgebra $su(2/1)$ these
normalization are fixed by the conditions (\ref{3-12}). Therefore
\begin{equation}
gA_{\mu}^{0}(x)\tilde{\lambda}_{0}=g'B_{\mu}(x)\lambda_{0},
\end{equation}
where the gauge field $B_{\mu}(x)$ is identified with $A_{\mu}^{0}(x)$. From this it follows
that (as opposed to the standard grand unification formula)
\begin{equation}\label{4-05}
g^{2}=\frac{10}{3}g'^{2}
\end{equation}
at the unification scale $M_{0}$. Using the standard formulas
$\sin^{2}\theta_{W}=g'^{2}/(g^{2}+g'^{2})$, $M_{H}^{2}=2\lambda v^{2}$ and
$M_{W}^{2}=g^{2}v^{2}/4$, we obtain from (\ref{2-03}) and (\ref{4-05}) the tree-level values
of the Weinberg angle and the Higgs boson mass
\begin{equation}
\sin^{2}\theta_{W}=\frac{3}{13},\quad M_{H}^{2}=\frac{8}{3}M_{W}^{2}.
\end{equation}
\par
Finally, we turn to the construction of the Lagrangian including fermion fields. Suppose
$\varPsi$ is a quartet with the fermions $\psi_{L}$, $\psi_{R_{1}}$ and $\psi_{R_{2}}$
transforming as a doublet together with two singlets in the internal space. Consider the
Lagrangian
\begin{equation}\label{4-03}
\mathcal{L}_{\varPsi}=\overline{\varPsi}i\gamma^{\mu}D_{\mu}\varPsi
-\left(k_{1}\overline{\varPsi}M\varPsi+k_{2}\overline{\varPsi}\gamma_{5}M\varPsi+h.c.\right),
\end{equation}
where the covariant derivative and the scalar field are defined in (\ref{3-05}) and
(\ref{3-06}). Substituting (\ref{3-08}) and (\ref{3-09}) into (\ref{4-03}), we get the
Lagrangian
\begin{align}\label{4-04}
\mathcal{L}^{(n)}_{\varPsi}&=\bar\psi_{L}i\gamma^{\mu}D_{\mu}\psi_{L}+\bar
\psi_{R_{1,2}}i\gamma^{\mu}D_{\mu}\psi_{R_{1,2}}\notag\\
&+\left(f_{1}\bar\psi_{L}\phi\psi_{R_{1}}+f_{2}\bar\psi_{L}\tilde{\phi}\psi_{R_{2}}
+h.c.\right),
\end{align}
where the constants (the Yukawa couplings)
\begin{equation}
f_{1}=\sqrt{\frac{n-1}{2n}}gk_{1},\quad f_{2}=\sqrt{\frac{n+1}{2n}}gk_{2}.
\end{equation}
Again passing to the quark-lepton representation, we get instead of (\ref{4-04}) the
Lagrangian describing the fermion sector of the standard electroweak model. Note that this
representation automatically describes the quarks and leptons correctly as transforming as
doublets and singlets under $SU(2)\times U(1)_{Y}$ with the correct quark and lepton charges.
Note also that $f_{2}=0$ as $n=-1$. Hence, neutrinos in the model are massless (in the
tree-level, of course).

\section{Weinberg angle}

\begin{figure}[htb]
\centering
\includegraphics[scale=0.3,angle=270]{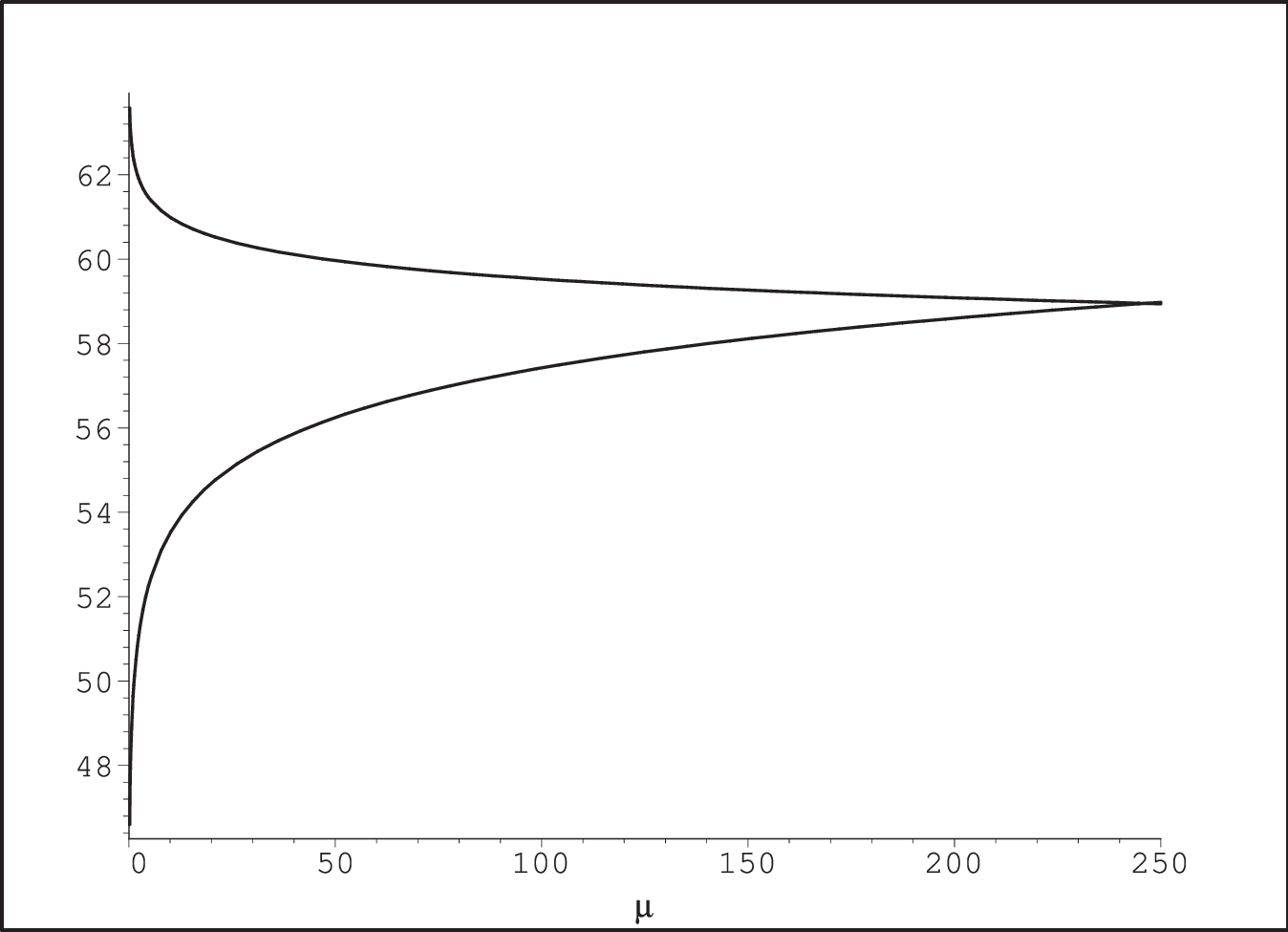}
\caption{\em Two-loop gauge coupling unification with $\alpha_{1}^{-1}(\mu)\geq
2\alpha_{2}^{-1}(\mu)$}.\label{fig1}
\end{figure}

The conditions (\ref{4-05}) are valid for the energy scale $\mu\geq M_{0}$. Now we study the
regime $\mu<M_0$. The evolution of the $SU(n)$ gauge coupling constant controlled by the
renormalization group equation
\begin{equation}\label{5-01}
\frac{d\alpha_{n}^{-1}(\mu)}{d\ln\mu}=\frac{b_{n}}{6\pi}.
\end{equation}
(For the time being we shall have ignored the contribution coming from the Higgs scalar and
higher-order effects.) It is important to keep in mind that the normalization used here is
non-canonical (as opposed to the grand unification model normalization). Because of the
different normalization of the $SU(2)$ generators, the gauge coupling constants of these two
models have different rates of evolution. For $SU(2)$, the value $b_{2}=22-2k^{-1}N_{f}$,
where $N_{f}$ is the number of quark flavors and $k=1$ (for the grand unification model) or
$2$ (in the non-canonical case). For $U(1)$, a straightforward summation of (squared) weak
hypercharges with proper rescaling of the normalization yields $b_1=-2N_{f}$ (in the both
cases). Expressing the low-energy couplings in terms of more familiar parameters, we can
represent the solutions of Eq. (\ref{5-01}) as
\begin{align}
\alpha^{-1}(\mu)\sin^2\theta_{\mu}
&=\alpha_{2}^{-1}(M_0)-\frac{b_2}{6\pi}\ln\frac{M_0}{\mu},\\
\frac35\alpha^{-1}(\mu)\cos^2\theta_{\mu}
&=\alpha_{1}^{-1}(M_0)-\frac{b_1}{6\pi}\ln\frac{M_0}{\mu},
\end{align}
where $\alpha_2(M_0)=k\alpha_{1}(M_0)$. Combining these equations, we easily obtain
\begin{equation}\label{5-02}
\sin^2\theta_{\mu}
=\frac{3}{5k+3}\left(1-\frac{55k\alpha(\mu)}{9\pi}\ln\frac{M_{0}}{\mu}\right).
\end{equation}
Using the mass relation
\begin{equation}
\sin^2\theta_{W}=1-\frac{M^2_{W}}{M^2_{Z}}
\end{equation}
and the experimental data (see~\cite{beri12})
\begin{align}
M_{Z}&=91.1876\pm 0,0021\,\,\text{GeV},\label{5-03}\\
M_{W}&=80.384\pm 0.015\,\,\text{GeV},\label{5-04}\\
\alpha^{-1}(M_{W})&=127.954\pm 0.015,\label{5-05}
\end{align}
we get $M_{0}\simeq 3\cdot 10^{13}$ GeV for $k=1$ (this is the standard unification scale in
GUT) and $M_{0}=246\pm10$ GeV for $k=2$. Thus for the non-canonical model, it can be argued
that the unification scale $M_{0}$ is coincided with the vacuum expectation value
\begin{equation}\label{5-06}
v=246.2204\pm0.0005\quad\mathrm{GeV}.
\end{equation}
Since our model do not contain new fields, this statement does not contradict the experimental
data. We show the gauge coupling unification in Figure~\ref{fig1}.

Conversely, substituting the values of $M_{Z}$, $\alpha(M_{W})$ and $M_{0}=v$ into
(\ref{5-02}), we easily obtain the value $M_{W}=80.3841\pm0.0018$ GeV. Taking into account the
contribution coming from the Higgs scalar $\delta^{(s)}M_{W}=-3.9$ MeV and two-order effects
(except for the gluons contribute) $\delta^{(2)}M_{W}=2.6$ MeV, we have
\begin{align}
M_{W}&=80.3835\pm0.0018\quad\mathrm{GeV},\\
\sin{\theta}_{W}&=0.22289\pm0.00005.
\end{align}
This excellent agreement with the experimental results. In order to account for the gluons
contribute, we need to change the normalization of the strong coupling. The appropriate
(non-canonical) normalization was proposed in~\cite{logi13}, where phenomenological
consequences of such normalization was investigated. In this case the contribution coming from
the two-order effects is $\delta^{(2)}M_{W}=1.9$ MeV that is not really affect the result.

\section{Higgs boson mass}

Our starting point in this section will be the effective potential of the SM. The effective
potential formalism of Coleman and Weinberg~\cite{cole73} and the functional improvements made
by Jackiw~\cite{jack74} are well known. In general one shifts scalar fields as
$\phi(x)\to\phi+\phi_{q}(x)$, where $\phi$ is $x$-independent. Then the effective potential is
given by the sum of vacuum graphs, with $\phi$-dependent propagators and vertices. At the tree
level the effective potential is $V_0(\phi)$, given by
\begin{equation}
V_0(\phi)=-\frac{1}{2}\mu^2\phi^2+\frac{1}{4}\lambda\phi^4,
\end{equation}
where $\lambda$ and $\mu^2$ are the SM scalar self-couplings. Writing the effective potential
as a sum
\begin{equation}
V_{eff}(\phi)=V_0(\phi)+V_1(\phi)+V_2(\phi)+\dots
\end{equation}
of the tree-level part $V_0(\phi)$ plus radiative corrections, one finds
\begin{equation}
M^2_{H}=\left.\frac{\partial^2 V_{eff}}{\partial\phi^2}\right|_{\phi=v}
\end{equation}
where $v$ is the vacuum expectation value at the minimum of the effective potential,
determined by the minimization condition
\begin{equation}\label{4-01}
\left.\frac{\partial V_{eff}}{\partial\phi}\right|_{\phi=v}=0.
\end{equation}
Then the mass term in the Higgs potential can be easily written as
\begin{equation}\label{6-02}
M_{H}^2=2\lambda v^2+\delta_{1}M_{H}^2+\delta_{2}M_{H}^2+\dots,
\end{equation}
where $\delta_{1}M_{H}^2$ and $\delta_{2}M_{H}^2$ are the one-loop and two-loop Higgs mass
corrections. In particular, it follows from (\ref{2-03}) and the standard formula $M_{W}=gv/2$
that in the tree-level the Higgs mass
\begin{equation}
M_{H}=\sqrt{\frac{8}{3}}M_{W}.
\end{equation}
Using the experimental knowledge of the vector boson mass, we find $M_{H}=131.267\pm0.023$
GeV.
\par
Now we turn our attention to the one-loop Higgs potential of the SM. It can be written in the
't Hooft-Landau gauge as~\cite{band93,ford93}
\begin{equation}\label{6-01}
V(\phi)=V_0(\phi)+V_1(\phi),
\end{equation}
where the one-loop Coleman-Weinberg potential is
\begin{align}\label{6-02}
V_1(\phi)&=\frac{1}{{16\pi^2}}\left[\frac{3}{2}W^2\left(\ln\frac{W}{M^2}
-\frac{5}{6}\right)\right.\notag\\
&+\frac{3}{4}Z^2\left(\ln\frac{Z}{M^2}-\frac{5}{6}\right)
+\frac{1}{4}H^2\left(\ln\frac{H}{M^2}-\frac{3}{2}\right)\notag\\
&+\frac{3}{4}G^2\left(\ln\frac{G}{M^2}-\frac{3}{2}\right)
\left.-3T^2\left(\ln\frac{T}{M^2}-\frac{3}{2}\right)\right]
\end{align}
with
\begin{align}
W&=\frac{1}{4}g^2\phi^2,\quad Z=\frac{1}{4}(g^2+g'^2)\phi^2\\
H&=-\mu^2+3\lambda\phi^2,\quad G=-\mu^2+\lambda\phi^2\\
T&=\frac{1}{2}g_{t}^2\phi^2.
\end{align}
Here $g$ and $g'$ are the gauge couplings and and $g_{t}$ is the top quark Yukawa coupling (we
neglect other Yukawa couplings throughout). At the minima of $V_0(\phi)$ we have $G=0$ and
$H$, $T$, $W$, $Z$ become the tree level (masses)$^{2}$ of the Higgs, top quark, $W$ and $Z$
bosons respectively. Using the minimum condition (\ref{4-01}), we can express $\mu^2$ in the
terms of $\lambda$ and $M$.
\par
\begin{figure}[htb]
\centering
\includegraphics[scale=0.3,angle=270]{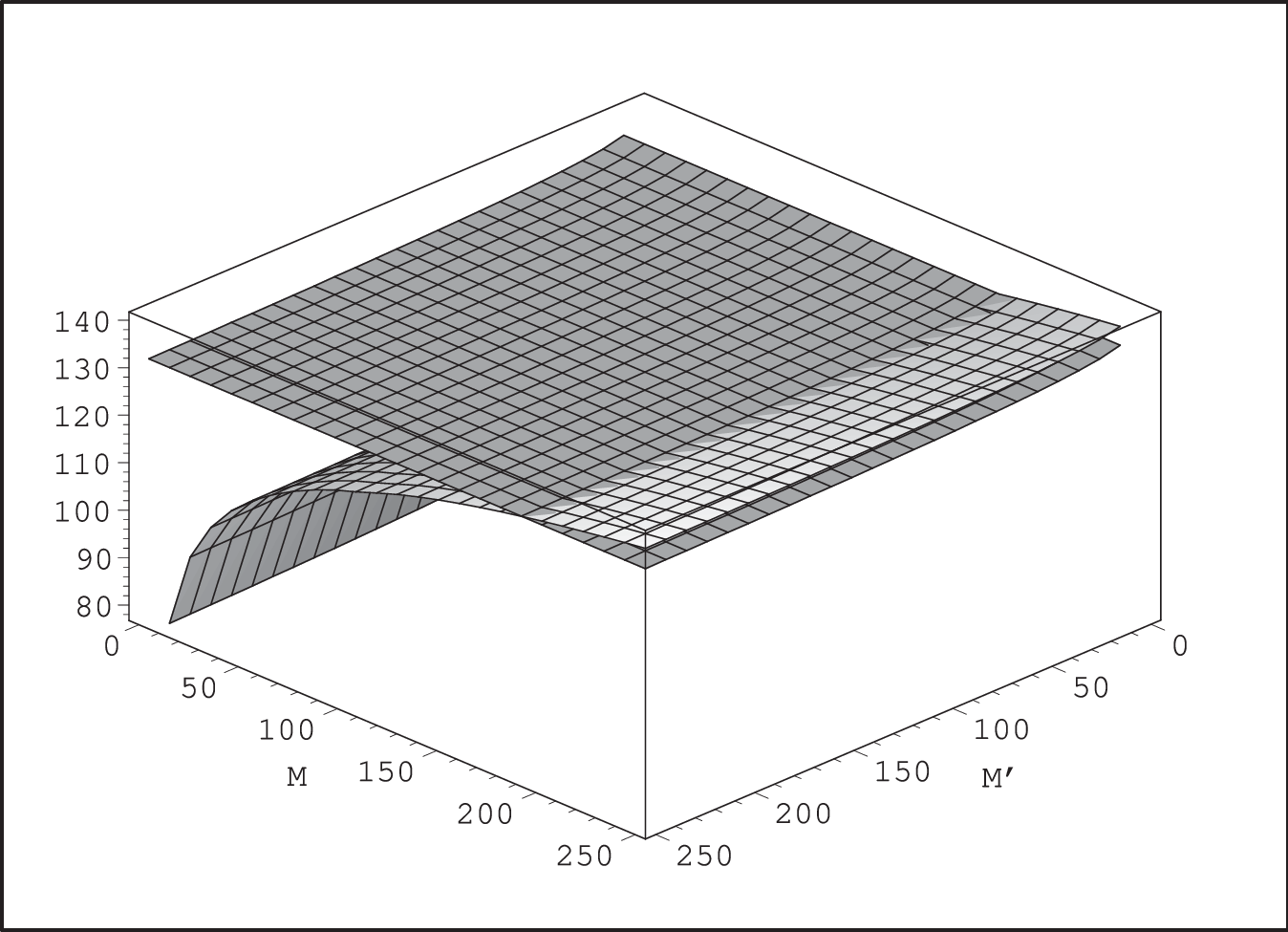}
\caption{\em Values of $M_{H}$ at the tree-level and one-loop approximations.}\label{fig2}
\end{figure}

On the other hand, it follows from (\ref{2-03}) that the running coupling
\begin{equation}\label{6-03}
\lambda^{-1}(M')=3\left[g^{-2}(M_{W})-\frac{b_2}{24\pi^2}\ln\frac{M_{W}}{M'}\right],
\end{equation}
where $b_{2}=22-2N_{F}-1/2$ (the number of quark flavours $N_{F}=5$ for a three-family theory
without the top quark). Note that the quartic coupling automatically inherit the good
ultra-violet asymptotically free behavior of the gauge coupling. Therefore we can present
$M_{H}$ as a function of the two parameters $M$ and $M'$. Using the standard formulas
\begin{equation}
g(M_{W})=\frac{\sqrt{4\pi\alpha(M_{W})}}{\sin\theta_{W}},\quad g_{t}=\frac{\sqrt{2}M_{T}}{v},
\end{equation}
and the experimental data (\ref{5-03})--(\ref{5-06}) and the top quark mass
$M_{T}=173.36\pm0.72$ GeV, that was computed in refs.~\cite{aalt12} and~\cite{chat12}, we find
this function (see Figure~\ref{fig2}).
\par
It was shown in~\cite{casa95} that the resulting minima and masses are relatively independent
of the precise choice of the mass parameter $M$, as long as the potential (\ref{6-02}) is used.
(Note that use of earlier potentials was inaccurate due to a sensitive dependence on the
choice of scale; see the review~\cite{sher00}.) Therefore we may suppose that it coincides
with $M'$. In this case, $M_{H}$ will be a function of only one parameter $M$ (see Figure~3).
\begin{figure}[htb]
\centering
\includegraphics[scale=0.3,angle=270]{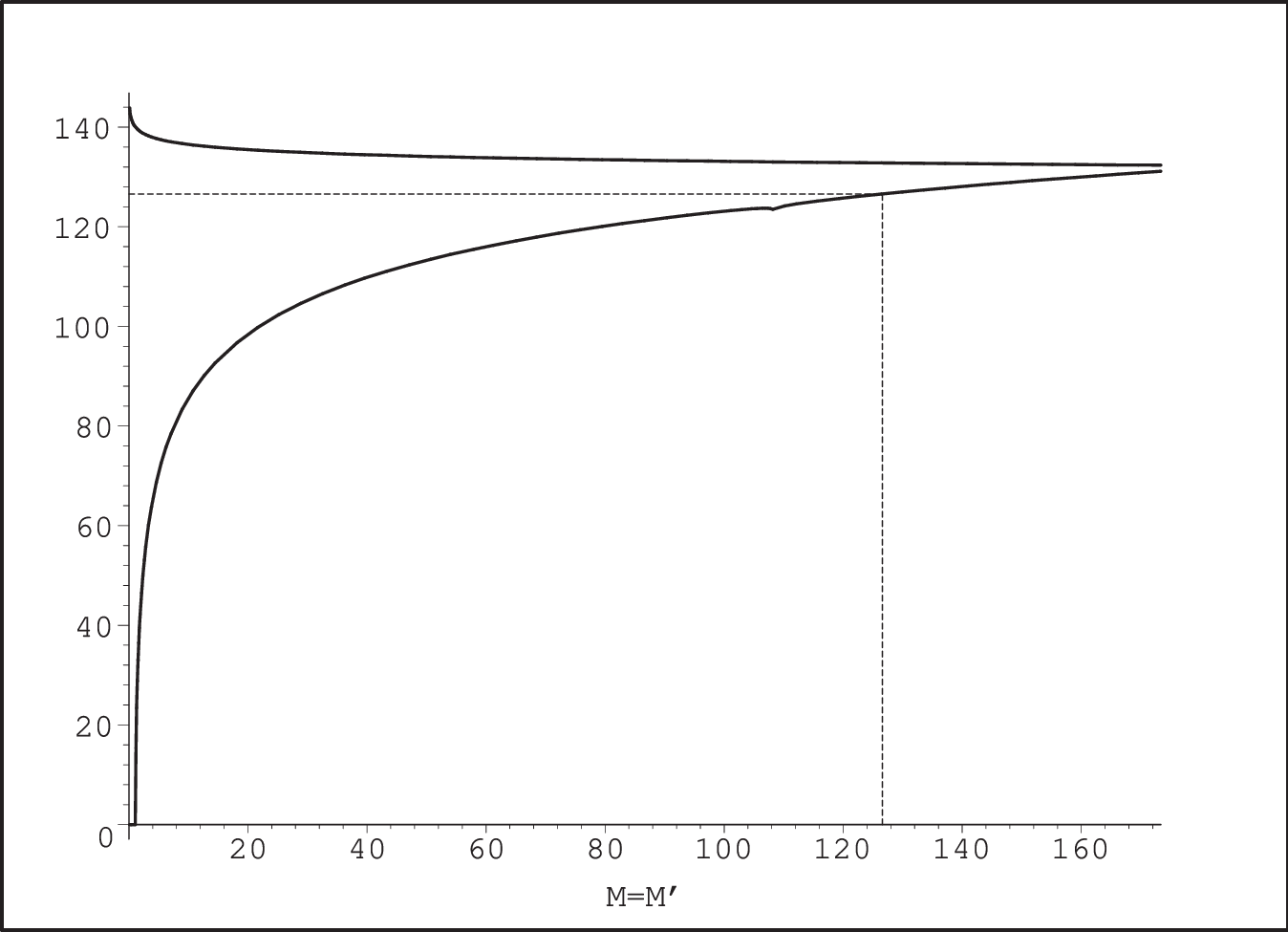}
\caption{\em Higgs boson mass at the tree-level and the one-loop (renormalised at $M=M'$).} \label{fig3}
\end{figure}
If we now take the value of $M$ at the production threshold of the Higgs boson we then obtain the Higgs mass in the one-loop approximation $M_{H}=126.58\pm 0.19$ GeV.
\par
Note that scalar loop contributions can be imaginary near the origin. However, this does not
destroy the calculational method, but is instead indicative of a physical instability. As was
shown in~\cite{wein87}, the imaginary part of the perturbatively calculated $V_{eff}$ has a
natural interpretation as the decay rate per unit volume corresponding to this process and
that it agrees quantitatively with an independent calculation of this rate. Since in the
considered case the imaginary contributions are very small, we can ignore them. The real
one-loop contributions to the Higgs mass are listed in Table~\ref{tab1}.
\begin{table}
$$
\begin{array}{c|ccccccc}
& m_{0} & \delta m_{V} & \delta m_{S} & \delta m_{F} & M_{H}\\ \hline
& 131.57 & -0.25 & -0.83 & -3.82 & 126.58 \\
\end{array}
$$
\caption{\em Values of the one-loop contribution to the Higgs mass. Here $\delta m_{V}$,
$\delta m_{S}$ and $\delta m_{F}$ are linearly approximate contributions of the vector and
scalar bosons and top quark respectively.}\label{tab1}
\end{table}
\par
In order that to obtain the two-loop approximation of the mass, one may use the results
of~\cite{ford92} (see also~\cite{degr12}), where the standard model effective potential to two
loops was calculated. But but it is easier to use a different approach. Let us write the mass
term in (\ref{6-02}) by
\begin{equation}
M_{H}^2=(M'_{H})^2+\delta_{2}M_{H}^2+\dots,
\end{equation}
where $M'_{H}$ is the mass term in the one-loop Higgs potential. Neglecting the three-loop
contribution to the Higgs mass and setting $\delta_{2}M_{H}=\delta_{2}M_{H}^{2}/2M_{H}$, one
obtains the approximate expression $M_{H}=M'_{H}+\delta_{2}M_{H}$. The value of the Higgs mass
$M_{H}$ renormalised at $M=M_{H}$ must coincide with the experimental value of the Higgs mass
measured by present ATLAS and CMS data~\cite{atla13,cms13} (see also average~\cite{giar14}):
\begin{equation}
M_{H}^{exp}=125.66\pm0.34\quad\mathrm{GeV}.
\end{equation}
Setting $M_{H}=125.66$ GeV and computing $M'_{H}$ (in the same normalization), we find the
value $\delta_{2}M_{H}=-0.81$ GeV. Adding this correction to the found previously value of the
one-loop Higgs mass, we obtain the Higgs mass in the two-loop approximation $M_{H}=125.77\pm
0.19$ GeV.
\begin{table}
$$\begin{array}{c|cccc}
& M & M_{H} & \lambda & m \\ \hline
\text{NLO} & 125.66 & 126.47 & 0.14547 & 132.81 \\
\text{NNLO} & 125.66 & 125.66 & 0.14377& 132.03 \\
\end{array}
$$
\caption{\em Values of the Higgs mass and the quartic coupling computed at one-loop and
two-loops and renormalised at $M=M_{H}^{exp}$ GeV.}\label{tab2}
\end{table}
\par
Note that the one-loop quartic Higgs coupling is fixed by the condition $M=M_{H}^{exp}$, while
the two-loop quartic coupling needs the condition $M_{H}=M$ instead of~(\ref{6-03}). It is
interesting that in both cases the tree-level equation $m^{2}=2\lambda v^2$ gives values of
the one-loop and two-loop Higgs mass terms computed at ref.~\cite{butt13}. We present these
calculations in Table~\ref{tab2}.

\section{Conclusion}

In this paper we addressed a question whether the observed Weinberg angle and Higgs mass are
calculable in the formalism based on a construction in which the electroweak gauge group
$SU(2)\times U(1)_{Y}$ is embedded in the graded Lie group $SU(2/1)$. The main result is that
the presented model predicts values of the Weinberg angle and the Higgs mass correctly up
to the two-loop level.
\par
In the paper we have followed the original works of Ne'eman and Fairlie believing that bosonic
fields of the model take their values in the superalgebra Lie $su(2/1)$ and fermionic
fields take their values in a representation space of $su(2/1)$. At the same time the model
contains a number of differences from the model of Ne'eman and Fairlie. The fundamental
difference is that while for them the gauge symmetry group is $SU(2/1)$, here we admit only
symmetries generated by its even subgroup, i.e., symmetries of the standard electroweak model.
Another difference is that we identify the Higgs potential with the extra component of a
five-dimensional vector field, whereas in the model of Ne'eman and Fairlie the six-dimensional
formalism are used. Finally, we are not limited only to the fundamental representation of
$SU(2/1)$. In the model we built a representation in which the one family fermions are
arranged in one multiplet.
\par
Interestingly, the model predicts the unification scale coinciding with the electroweak scale.
It fits perfectly with the choice of the symmetry group of the model, since this scale
characterizes the spontaneous symmetry breaking of $SU(2)\times U(1)_{Y}$ to $U(1)$. Since the
model do not contain new fields, it follows that the appearance of the unification scale does
not contradict the experimental data. Note also that the model automatically describes the
quarks and leptons correctly as transforming as doublets and singlets under the gauge group
with the correct quark and lepton charges and also predicts masslessness of the neutrinos.

\appendix
\section{Representations of $su(2/1)$}

The smallest nontrivial simple Lie superalgebra $su(2/1)$ contains (in the fundamental
representation) four bosonic generators $\lambda_{a}$ $(a\leq3)$ which form the Lie algebra
$su(2)\oplus u(1)$ and four fermionic generators $\lambda_{\bar{a}}$ $(\bar{a}>3)$, whose
commutation relations read as
\begin{align}
[\lambda_{a},\lambda_{b}]&=2ic_{abc}\lambda_{c},\label{A-01}\\
[\lambda_{a},\lambda_{\bar{b}}]&=2if^{\bar{c}}_{a\bar{b}}\lambda_{\bar{c}},\label{A-02}\\
\{\lambda_{\bar{a}},\lambda_{\bar{b}}\}&=2d^{c}_{\bar{a}\bar{b}}\lambda_{c}.\label{A-03}
\end{align}
Here $\lambda_{1},\dots,\lambda_{7}$ are the standard Gell-Mann matrices and
$\lambda_{0}=\text{diag}(-1,-1,-2)$. There is an irreducible four dimensional representation
of the $su(2/1)$ superalgebra. The existence of this representation is a simple consequence of
the isomorphism between $su(2/1)$ and $osp(2/2)$, which is a generalization of the well known
isomorphisms between the first members of the infinite families of simple Lie algebras. The
four bosonic lambda matrices read:

{\small\begin{equation} \lambda_{k}=\begin{pmatrix}\sigma_{k}&0\\0&0\end{pmatrix},\quad
\lambda_{0}
=\frac{1}{n}\begin{pmatrix}1&0&0&0\\0&1&0&0\\0&0&1+n&0\\0&0&0&1-n\end{pmatrix},\notag
\end{equation}}

\noindent where $\sigma_{k}$ are the standard Pauli matrices and $n$ is a nonzero real number.
The four fermionic lambda matrices read:

{\small\begin{equation}
\lambda_{4}=\frac{1}{\sqrt{2n}}\begin{pmatrix}0&0&0&\sqrt{n-1}\\0&0&\sqrt{n+1}&0
\\0&-\sqrt{n+1}&0&0\\\sqrt{n-1}&0&0&0\end{pmatrix},\notag\\
\end{equation}
\begin{equation}
\lambda_{5}=\frac{i}{\sqrt{2n}}\begin{pmatrix}0&0&0&-\sqrt{n-1}\\0&0&\sqrt{n+1}&0
\\0&\sqrt{n+1}&0&0\\\sqrt{n-1}&0&0&0\end{pmatrix},\notag\\
\end{equation}
\begin{equation}
\lambda_{6}=\frac{1}{\sqrt{2n}}\begin{pmatrix}0&0&-\sqrt{n+1}&0\\0&0&0&\sqrt{n-1}
\\\sqrt{n+1}&0&0&0\\0&\sqrt{n-1}&0&0\end{pmatrix},\notag\\
\end{equation}
\begin{equation}
\lambda_{7}=\frac{-i}{\sqrt{2n}}\begin{pmatrix}0&0&\sqrt{n+1}&0\\0&0&0&\sqrt{n-1}
\\\sqrt{n+1}&0&0&0\\0&-\sqrt{n-1}&0&0\end{pmatrix}.\notag
\end{equation}}

\noindent Note that as $n=-1$, we obtain the representation identical to the fundamental
representation of $su(1/2)$.

\end{document}